\documentclass[pra,twocolumn,tightenlines,showpacs,nofootinbib]{revtex4}
\usepackage{bm,dcolumn,amsmath,graphicx}

\newcommand{\etal}{{\it et al.}}
\newcommand{\eref}[1]{Eq.~(\ref{#1})}
\newcommand{\tref}[1]{Table~\ref{#1}}

\begin{document}

\title{Transition frequency shifts with fine-structure
constant variation for {Fe~I}. Isotope shift calculations
in {Fe~I} and {Fe~II}}

\author{S. G. Porsev$^{1,2}$}
\author{M. G. Kozlov$^{1,2}$}
\author{D. Reimers$^2$}
\affiliation{$^1$ Petersburg Nuclear Physics Institute, Leningrad district,
Gatchina, 188300, Russia}
\affiliation{$^2$~Hamburger Sternwarte, Universit\"{a}t Hamburg, Hamburg, Germany}

\date{ \today }

\begin{abstract}
In this paper we calculated
the relativistic corrections to transition frequencies ($q$ factors)
of Fe~\textsc{i} for the transitions from the even- and odd-parity states
to the ground state. We also carried out isotope shift calculations
in Fe~\textsc{i} and Fe~\textsc{ii}.
To the best of our knowledge,
the calculation of the IS in Fe~\textsc{i} was performed for the first time.
\end{abstract}

\maketitle

\section{Introduction}
\label{sec_intro}
The problem of temporal and spacial variation of the fundamental
physical constants is actively discussed in the literature
during last several years. A recent review of its current status can be
found elsewhere~\cite{GarIseKub07}. One of the reasons stimulating
this activity was the discovery of acceleration of the
universe (for a review see~\cite{CopSamTsu06}), what is usually associated with the
existence of the dark energy. Modern theories describing
cosmological evolution predict that the dark energy may cause
variations of the fundamental constants.

A statement that the fine-structure constant $\alpha$ has possibly
changed during evolution of the universe was made by Australian
group in Ref.~\cite{MurWebFla03}. Other astrophysical groups do not
confirm this result~\cite{QRL04,SCP04},
and hence new laboratory and astrophysical investigations are
required.

Laboratory studies of hypothetical variation of the fine-structure
constant are based on the fact that
transition frequencies in atoms depend on $\alpha Z$, where $Z$ is atomic number.
Supposing that the nowadays value of $\alpha$ differs from its value in the earlier
universe we can study space-time variation of $\alpha$ by
comparing atomic frequencies for distant objects in the universe
with their laboratory values. In practice, we need to find relativistic
frequencies shifts, determined by so-called $q$ factors, according to
\begin{align}
\omega = \omega_\mathrm{lab} + q x, \quad x \equiv
\left({\alpha}/{\alpha_\mathrm{lab}}\right)^2 - 1\,.
\label{qfactor1}
\end{align}

Most advantageous for these studies are the atoms and ions
for which $q$ factors of transitions between certain states
significantly differ from each other.
At the same time these elements should be abundant in
the universe to provide sufficient observable data.
Fe~\textsc{i} and Fe~\textsc{ii} discussed in this paper meet
both these conditions.
Spectra of Fe~\textsc{ii} were used by several groups
\cite{MurWebFla03,QRL04,SCP04} within ``many-multiplet method'' and
by Levshakov \etal\ within ``single ion differential alpha
measurement method'' \cite{LML07}. The atoms of Fe~\textsc{i}
were observed in resonance ultraviolet lines in two damped Ly$\alpha$
systems at $z$=0.452~\cite{Dod07} and $z$=1.15~\cite{QuaReiBaa08}.
According to~\cite{ML07} the
spectra of Fe~\textsc{i} are also observed for the high redshift
quasars and may be used for the $\alpha$-variation search.
In this work we calculate the $q$ factors for the transitions in
Fe~\textsc{i} from the even- and odd-parity states to the ground state.

As was pointed out in many papers
including~\cite{Lev94,VPI01,MurWebFla03,KozKorBer04,KorKoz07} one
of the problems that occurs in the study of possible $\alpha$
variation is a necessity to separate this effect from the isotope
shift (IS) effect. A method to resolve this problem was suggested
in~\cite{KozKorBer04}. This method requires (along with a knowledge
of atomic relativistic coefficients $q$) precise calculation of the
isotope shift coefficients.

In this paper we carry out isotope shift calculations in Fe~\textsc{i}
and Fe~\textsc{ii}. We compare the results obtained for Fe~\textsc{ii} with
other available data. To the best of our knowledge, the calculation of the IS
in Fe~\textsc{i} is performed for the first time.

The paper is organized as follows.
Sec.~\ref{calc_FeI} is devoted to the method of calculation of the
properties of Fe~\textsc{i}. We discuss the results obtained for the
$q$ factors. In Sec.~\ref{ms_in_FeI_and_II} the method of calculation of isotope
shift is described. We present the results of the isotope shift
calculation in Fe~\textsc{i} and Fe~\textsc{ii}.
Finally, Sec.~\ref{conclusions} contains concluding remarks. Atomic
units ($\hbar = |e| = m_e = 1$) are used throughout the paper.

\section{Calculation of $q$ factors for ${\rm \bf Fe~I}$}
\label{calc_FeI}

\subsection{Method of calculation}
\label{sec_method}

To find $q$ factors we need to solve the atomic relativistic
eigenvalue problem for different values of $\alpha$ or,
respectively, for different values of $x$ from~\eref{qfactor1}.
We can calculate atomic frequencies $\omega_\pm$ for two values $x=\pm
1/8$ of the parameter $x$.
Our experience shows that such a choice of $x$ allows us to
meet two conditions. The value $|x|=1/8$ is usually sufficiently small to neglect
nonlinear corrections and sufficiently large to make calculations
numerically stable.
The corresponding $q$ factor is given by
\begin{align}
q = 4 (\omega_+ - \omega_-) .
\label{qfactor2}
\end{align}

The ground state configuration of Fe~\textsc{i}
is ($1s^2 ...\, 3d^6 4s^2$). Since it has eight
electrons in open shells its spectrum is rather dense and
complicated. Due to proximity of the levels with the
same total angular momentum (especially in
the middle of spectrum which astrophysically is most
interesting) they strongly interact with each other.
All this makes calculations of Fe~\textsc{i} very difficult.
To the best of our knowledge the only calculation of
$q$ factors for neutral iron was carried out recently by Dzuba and
Flambaum in Ref.~\cite{DzuFla08}. They used a simple method
combining {\it ab initio} Hartree-Fock and configuration interaction (CI)
technique with some semiempirical fitting of energy levels.

In this paper we make pure {\it ab initio} calculations in the
frame of the eight-electron CI method. The [$1s^2 ...\, 3p^6$]
electrons are treated as core electrons while $3d$, $4s$, and $4p$ electrons
are in the valence space.
The number of configurations accounted for in our calculations is noticeably greater
than in~\cite{DzuFla08}. As a result an effect of configuration interaction
is treated more accurately. Below we will discuss it more detailed.

We started from solving the Dirac-Hartree-Fock equations. The self-consistency
procedure was done for ($1s^2 ...\, 3d^6 4s^2$) configuration. After that
the $4p_{1/2}$ and $4p_{3/2}$ orbitals were constructed as follows. All
electrons were frozen; one electron from the $4s$ shell was moved to the $3d$ shell
and another electron from the $4s$ shell was moved to $4p$ shell.
Thus, the valence orbitals $4p_{1/2}$ and $4p_{3/2}$ were constructed
for the $3d^7 4p$ configuration.

On the next stage we constructed virtual orbitals. We used the method
described in~\cite{Bog91,KozPorFla96} and applied by us
for calculating different properties of Fe~\textsc{ii}~\cite{PorKosTup07}.
In this method an upper component of virtual
orbitals is formed from the previous orbital of the same symmetry by
multiplication by some smooth function of radial variable $r$. The
lower component is then formed using kinetic balance condition.

Our basis sets included $s,p,d,$ and $f$ orbitals with principle
quantum number $n\le N$. We designate them as $[N\!spdf]$.
We have carried out the calculations of energy levels, $g$ and $q$ factors for
three basis sets with $N=4\div6$.
Configuration space was formed by single and double (SD) excitations
from the configurations $3d^6 4s^2$, $3d^6 4s\,4p$, and $3d^7 4p$.

Additionally we studied Breit corrections including the Breit interaction
into consideration. For $4spdf$ basis set we computed
$q$ factors in the Coulomb-Breit approximation and compared them with
the results obtained in pure Coulomb approximation. As analysis showed
for a majority of states the Breit interaction changed the values of
the $q$ factors only at the level of few percent. For this reason
all results which we discuss below are obtained in the pure
Coulomb approximation.

\subsection{Ground multiplet $^5\!D_J$}
\label{Fe:GS}

The ground state FS transitions in mid- and far-infrared were observed
in emission for different redshifts for a number of atoms and ions
(see,e.g.,\cite{GreDyaGeb77,MooSalFur80,WeiHenDow03}).
The infrared FS lines of the neutral
iron have not been detected yet in astronomical objects.
But a such detection is expected in extragalactic objects at a new
generation of telescopes like the Stratospheric Observatory for
Infrared Astronomy and Far Infrared and Submillimeter Telescope.

We start discussion from the results obtained for transitions
between the states of the ground multiplet.
The CI space corresponds to SD
excitations from the configuration $3d^6 4s^2$. In~\tref{Tab_ev_wgq}
we present results obtained for the [$6spdf$] basis set
for energy levels, $g$ factors, and $q$ factors for transitions from the
ground state $^5\!D_4$.
\begin{table}[bt]
\caption{Energy levels (cm$^{-1}$), $g$ factors, and $q$ factors (cm$^{-1}$)
for the ($3d^6 4s^2$)  $a\,^5\!D_J$ states.
The values are obtained for the [$6spdf$] basis set.}

\label{Tab_ev_wgq}

\begin{ruledtabular}
\begin{tabular}{lccccc}
&\multicolumn{2}{c}{Experiment\footnotemark[1]}
&\multicolumn{3}{c}{Calculations} \\
& \multicolumn{1}{c}{Energy} & \multicolumn{1}{c}{$g$}
& \multicolumn{1}{c}{Energy} & \multicolumn{1}{c}{$g$} & \multicolumn{1}{c}{$q$} \\[1mm]
\hline
$a\,^5\!D_4$   &     0 & 1.5002  &     0 & 1.4996  &     0 \\
$a\,^5\!D_3$   &   416 & 1.5003  &   407 & 1.4994  &   413 \\
$a\,^5\!D_2$   &   704 & 1.5004  &   694 & 1.5001  &   684 \\
$a\,^5\!D_1$   &   888 & 1.5002  &   879 & 1.4999  &   853 \\
$a\,^5\!D_0$   &   978 &         &   969 &         &   935
\end{tabular}
\end{ruledtabular}
\footnotemark[1]{NIST, Ref.~\cite{NIST}.}
\end{table}

A method to use these fine-structure (FS) transitions to study
$\alpha$ variation at very high redshifts was suggested in Ref.~\cite{KozPorLev08}.
This method crucially depends on the differences of the dimensionless
sensitivity coefficients defined as
\begin{equation}
Q_{J,J-1} = (q_J - q_{J-1})/\omega_{J,J-1}, \label{Eq:Q}
\end{equation}
where $\omega_{J,J-1} = E_J - E_{J-1}$ is the frequency of the FS
transition $J \leftrightarrow J-1$.

It was shown in~\cite{KozPorLev08} that for the levels of the
$^{2S+1}\!L_J$ multiplet the difference, $\Delta Q$, between the
dimensionless sensitivity coefficients $Q_{J,J-1}$ and $Q_{J-1,J-2}$
is given by the following formula
 \begin{equation}
 \Delta Q \equiv Q_{J,J-1}-Q_{J-1,J-2}
 = \frac{J-1}{J} \left(
 \frac{\omega_{J,J-1}}{\omega_{J-1,J-2}} \right) - 1,
 \label{Eq:DelQ}
 \end{equation}
which links $\Delta Q$ to the experimentally observed FS transition
frequencies. This analytical expression is valid up to the terms
of the order of $(\alpha Z)^4$.
Note, that in the first order in the
spin-orbit interaction the left hand side of \eref{Eq:DelQ} turns to
zero. For this reason $\Delta Q$ is very sensitive to the ratio of the
FS transition frequencies.

As is seen from~\tref{Tab_ev_wgq} we reproduce the FS intervals with
a few percent accuracy. The uncertainties of the $q$ factors
listed in~\tref{Tab_ev_wgq} can also be estimated at the level of a few
percent. Such inaccuracies in calculation
of the FS transition frequencies and $q$ factors, though small, are sufficient to result in significant
differences between the values of $\Delta Q$ calculated with use of~\eref{Eq:Q} and
obtained from~\eref{Eq:DelQ}.

To illustrate it we present in~\tref{Tab_DelQ} the differences of the sensitivity
coefficients ($\Delta Q$) of the FS lines within the ground multiplet
$a ^5\!D_J$ obtained by both methods mentioned above. The
corresponding entries in~\tref{Tab_DelQ} are denoted as~\eref{Eq:Q}
and~\eref{Eq:DelQ}. In the first case we used the $q$ factors and the calculated transition
frequencies from \tref{Tab_ev_wgq}. In the second case the~\eref{Eq:DelQ} and the
experimental frequencies were used.
\begin{table}[bt]
\caption{$\Delta Q$ for the FS lines within the ground multiplet $a
^5\!D_J$. Numerical results are calculated using \eref{Eq:Q} for the
basis set [$6\!spdf$]. Analytical expression \eqref{Eq:DelQ} is
applied to the experimental frequencies as in Ref.~\cite{KozPorLev08}.}

\label{Tab_DelQ}

\begin{ruledtabular}
\begin{tabular}{cccc}
\multicolumn{1}{c}{($J_a, J'_a$)} & \multicolumn{1}{c}{($J_b, J'_b$)}
& \eref{Eq:Q} & \eref{Eq:DelQ}  \\[1mm]
\hline
(2,3)  &  (3,4)  & 0.067 &  0.083  \\
(1,2)  &  (2,3)  & 0.027 &  0.043  \\
(0,1)  &  (1,2)  & 0.018 &  0.024  \\
\end{tabular}
\end{ruledtabular}
\end{table}

Comparing these results we see, that there is only qualitative agreement between numerical
and analytical approaches. For instance, \eref{Eq:DelQ} predicts decreasing of $\Delta Q$
with decreasing $J$. When we use~\eref{Eq:Q}
we observe the same behavior. At the same time, quantitatively, a disagreement
is rather significant. We think that numerical errors in computing the $q$ factors
and the FS transition frequencies are more essential for calculation of $\Delta Q$
than the terms $\sim (\alpha Z)^6$ (and higher) neglected in~\eref{Eq:DelQ}. For this reason
we consider the values obtained from~\eref{Eq:DelQ} as more correct.

\subsection{Odd parity levels}
\label{Fe:results}
The transitions from the ground state to the odd-parity states are observed
in absorbtion in the spectra of quasars.
The results of the eight-electron CI calculations of the energy
levels, $g$ factors and $q$ factors for the odd-parity states of the $3d^6
4s4p$ configuration of Fe~\textsc{i} are listed
in~\tref{Tab_od_wgq}.
We have carried out calculations for the three
basis sets [$(4\div6)spdf$]. The CI space for each basis set
corresponds to SD excitations from two configurations $3d^6 4s4p$
and $3d^7 4p$. The agreement between theoretical and experimental
energy levels was systematically improved  with increasing the basis
set and best results were obtained for the longest [$6spdf$] basis
set.
In~\tref{Tab_od_wgq} we present the results obtained for the largest [$6spdf$] basis set.
We restrict ourselves by consideration of the states with total angular momenta $J$=3,4, and, 5
because only to these states there are strong electric dipole transitions
from the ground  ($3d^6 4s^2 \,\,  ^5\!D_4$) state.
\begin{table}[bt]
\caption{Energy levels (cm$^{-1}$), $g$ factors, and $q$ factors (cm$^{-1}$)
for the odd-parity energy levels of the $3d^6 4s4p$ configuration.
The values are obtained for the [$6spdf$] basis set.}

\label{Tab_od_wgq}

\begin{ruledtabular}
\begin{tabular}{lccccrr}
& \multicolumn{2}{c}{Experiment\footnotemark[1]} & \multicolumn{4}{c}{Calculations} \\
& \multicolumn{1}{c}{Energy} & \multicolumn{1}{c}{$g$}
& \multicolumn{1}{c}{Energy} & \multicolumn{1}{c}{$g$} & \multicolumn{1}{c}{$q$}
& \multicolumn{1}{c}{$q$~\cite{DzuFla08}} \\[1mm]
\hline
$z\,^7\!D^o_5$ & 19351 & 1.597   & 20204 & 1.599 &   722   &   490    \\
$z\,^7\!D^o_4$ & 19562 & 1.642   & 20415 & 1.649 &   895   &   662    \\
$z\,^7\!D^o_3$ & 19757 & 1.746   & 20613 & 1.749 &  1096   &   891    \\[0.8mm]
$z\,^7\!F^o_5$ & 22846 & 1.498   & 22988 & 1.500 &  1004   &   827    \\
$z\,^7\!F^o_4$ & 22997 & 1.493   & 23148 & 1.501 &  1155   &   982    \\
$z\,^7\!F^o_3$ & 23111 & 1.513   & 23269 & 1.501 &  1269   &  1103   \\[0.8mm]
$z\,^7\!P^o_4$ & 23711 & 1.747   & 23903 & 1.749 &   744   &   491    \\
$z\,^7\!P^o_3$ & 24181 & 1.908   & 24383 & 1.915 &  1211   &   983    \\[0.8mm]
$z\,^5\!D^o_4$\footnotemark[2]
               & 25900 & 1.502   & 26065 & 1.497 &  1032   &   999    \\
$z\,^5\!D^o_3$ & 26140 & 1.500   & 26302 & 1.496 &  1239   &  1223   \\[0.8mm]
$z\,^5\!F^o_5$\footnotemark[2]
               & 26875 & 1.399   & 26820 & 1.400 &   927   &   880    \\
$z\,^5\!F^o_4$ & 27167 & 1.355   & 27114 & 1.353 &  1224   &  1180   \\
$z\,^5\!F^o_3$ & 27395 & 1.250   & 27338 & 1.255 &  1441   &  1402   \\[0.8mm]
$z\,^5\!P^o_3$\footnotemark[2]
               & 29056 & 1.657   & 28653 & 1.665 &  1008   &   859    \\[0.8mm]
$z\,^3\!F^o_4$ & 31307 & 1.250   & 31216 & 1.250 &  1177   &  1267   \\
$z\,^3\!F^o_3$ & 31805 & 1.086   & 31703 & 1.093 &  1665   &  1808   \\[0.8mm]
$z\,^3\!D^o_3$ & 31323 & 1.321   & 31177 & 1.324 &  1338   &  1456   \\[0.8mm]
$y\,^5\!P^o_3$ & 36767 & 1.661   & 38411 & 1.665 &   910   &         \\[0.8mm]
$x\,^5\!D^o_4$\footnotemark[2]
               & 39626 & 1.489   & 41328 & 1.499 &  2163   &  1680   \\
$x\,^5\!D^o_3$ & 39970 & 1.504   & 41753 & 1.494 &  2632   &         \\[0.8mm]
$x\,^5\!F^o_5$\footnotemark[2]
               & 40257 & 1.390   & 40476 & 1.399 &  1725   &  1042   \\
$x\,^5\!F^o_4$ & 40594 & 1.328   & 40839 & 1.347 &  2158   &         \\
$x\,^5\!F^o_3$ & 40842 & 1.254   & 41077 & 1.254 &  2416   &         \\
\end{tabular}
\end{ruledtabular}
\footnotemark[1]{NIST, Ref.~\cite{NIST}}; \\
\footnotemark[2]{States observed in the quasar absorption spectra}.
\end{table}

Comparing the $q$ factors found in this work with the results obtained in Ref.~\cite{DzuFla08} we
see reasonable agreement between them for majority of the states. At the
same time some differences reach 40\%.
One of the reason of these discrepancies is the strong configuration
interaction for ceratin levels that significantly influences on the
$q$ factors. For instance for the astrophysically interesting
$x\,^5\!D^o_4$ and $x\,^5\!F^o_5$ states of the $3d^6 4s4p$
configuration we obtained in one-configurational approximation the
following $q$ factors: $q(x\,^5\!D^o_4)$ = 835 cm$^{-1}$ and
$q(x\,^5\!F^o_5)$ = 861 cm$^{-1}$. As is seen from~\tref{Tab_od_wgq}
these values are more than two times smaller than those obtained for
the [$6spdf$] basis set, when a large number of configurations was
included into consideration. At the same time we see that even for
the largest basis set the number of configurations taken into
account is still not sufficient to correctly reproduce the order of
certain states. For instance, the energy levels of the
$x\,^5\!D^o_J$ multiplet lays higher than the energy levels of the
$x\,^5\!F^o_J$ multiplet. Since this range of spectrum is very dense
and there are many nearby levels with the same total angular momenta
it can lead to incorrect interaction of the mentioned states with
their neighbours.

Now we will discuss the odd-parity states of the $3d^7 4p$ configuration.
Correct calculation of different properties of these states is more difficult
than calculation of the states belonging to the $3d^6\,4s4p$ configuration.
First, the states of the $3d^7 4p$ configuration are located higher in energy
than the majority of the states of the $3d^6\,4s4p$ configuration considered by us.
Second, the energy levels of the $3d^7 4p$ configuration belonging to different multiplets
are located very close to each other and, respectively, strongly interact to each other.
An additional problem is that the HFD equations were solved self-consistently for the
configuration $3d^6\,4s^2$. The configuration $3d^7\,4p$ differs more significantly
from it than the configuration $3d^6\,4s4p$, in particular, by the presence of an extra
$d$ electron on the $3d$ shell. As a consequence, it is more difficult to reproduce
the correct transition frequencies from the states of the $3d^7\,4p$ configuration
to the ground state. As our analysis shows these calculated frequencies tend to be
larger than the experimental ones.

To investigate how different properties of these states will change when
the basis set is increased from [$4spdf$] to [$6spdf$] we have carried out
the calculations for all these basis sets. The agreement between theoretical and
experimental frequencies was at the level of 5-10\%. As it turned out the
best agreement was achieved for the [$5spdf$] basis set.
This obviously indicates that for this configuration we are still far
from saturating the CI space.
In~\tref{Tab_d7p} we present energy levels, $g$ and $q$ factors
obtained for this basis set.

\begin{table}[bt]
\caption{Energy levels (cm$^{-1}$), $g$ factors,
and $q$ factors (cm$^{-1}$) for the states of the $3d^7 4p$ configuration.
The values are obtained for the $5spdf$ basis set.
Results obtained in Ref.~\cite{DzuFla08} are given for comparison}

\label{Tab_d7p}

\begin{ruledtabular}
\begin{tabular}{lccccrr}
& \multicolumn{2}{c}{Experiment\footnotemark[1]} & \multicolumn{4}{c}{Calculations} \\
& \multicolumn{1}{c}{Energy} & \multicolumn{1}{c}{$g$}
& \multicolumn{1}{c}{Energy} & \multicolumn{1}{c}{$g$} & \multicolumn{1}{c}{$q$}
& \multicolumn{1}{c}{$q$~\cite{DzuFla08}} \\[1mm]
\hline
$y\,^5\!D^o_4$ & 33096 & 1.496   & 34766 & 1.499    & 1794   & 2494 \\
$y\,^5\!D^o_3$ & 33507 & 1.492   & 35140 & 1.500    & 2118   & 3019 \\[0.8mm]
$y\,^5\!F^o_5$ & 33695 & 1.417   & 35750 & 1.399    & 2306   & 2672 \\
$y\,^5\!F^o_4$ & 34040 & 1.344   & 36069 & 1.348    & 2585   & 3021 \\
$y\,^5\!F^o_3$ & 34329 & 1.244   & 36338 & 1.248    & 2836   & 3317 \\[0.8mm]
$z\,^5\!G^o_5$ & 34782 & 1.218   & 37164 & 1.212    & 2736   & 3024 \\
$z\,^5\!G^o_4$ & 35257 & 1.103   & 37646 & 1.088    & 3196   & 3520 \\
$z\,^5\!G^o_3$ & 35612 & 0.887   & 38038 & 0.855    & 3563   & 3864 \\[0.8mm]
$z\,^3\!G^o_5$ & 35379 & 1.248   & 37806 & 1.256    & 3343   & 3340 \\
$z\,^3\!G^o_4$ & 35768 & 1.100   & 38142 & 1.120    & 3640   & 3697 \\
$z\,^3\!G^o_3$ & 36079 & 0.791   & 38401 & 0.821    & 3909   & 4096 \\[0.8mm]
$y\,^3\!F^o_4$ & 36686 & 1.246   & 38703 & 1.245    & 3083   & 3085 \\
$y\,^3\!F^o_3$ & 37162 & 1.086   & 39155 & 1.091    & 3447   & 3487 \\[0.8mm]
$y\,^3\!D^o_3$ & 38175 & 1.324   & 39904 & 1.321    & 3079   &      \\[0.6mm]
\end{tabular}
\end{ruledtabular}
\footnotemark[1]{NIST, Ref.~\cite{NIST}.}
\end{table}

As is seen from~\tref{Tab_d7p} the agreement between theoretical and
experimental energy levels is worse than it was for the states of
the $3d^6 4s4p$ configuration but nevertheless the largest difference
does not exceed 7\%. An interesting fact is that the agreement
between $q$ factors obtained by us and by Dzuba and Flambaum~\cite{DzuFla08}
is (on average) better than for the states of the $3d^6 4s4p$ configuration.

Analyzing the $q$ factors presented in Tables~\ref{Tab_od_wgq} and \ref{Tab_d7p}
we see that the former are (on average) smaller than the latter. It has simple
explanation because the transition between $3d^6 4s^2$ and $3d^6 4s4p$ configurations
is basically a  one-electron $4s$-$4p$ transition. The transition between $3d^6 4s^2$
and $3d^7 4p$ configurations is due to simultaneous $4s$-$4p$ and $4s$-$3d$  one-electron
transitions. As was shown in~\cite{DzuFlaWeb99}, when $\alpha$ is changing
towards its nonrelativistic limit $\alpha=0$, changes of
one-electron energies of $s$ and $d$ states are differently
directed. It leads to increasing the $q$ factors.

Note that the states nominally belonging to the $3d^7 4p$
configuration are formed, as a rule, as a result of strong mixing of $3d^6 4s4p$ and
$3d^7 4p$ configurations. An admixture of the $3d^6 4s4p$ configuration to the
$3d^7 4p$ configuration leads to opening a strong electric dipole transitions from the states
of the $3d^7 4p$ configuration to the ground state. It allows one to make these transitions
observable.

Using the results obtained for the [$4spdf$], [$5spdf$], and [$6spdf$] basis sets
we are able to estimate uncertainties of the calculated $q$ factors.
We roughly estimate the uncertainty as the difference between largest and
smallest values of the $q$ factors found for the three basis sets mentioned above.
Such conservative estimate allows us to say that the $q$ factors for the odd-parity
energy levels listed in Tables~\ref{Tab_od_wgq} and \ref{Tab_d7p} are within 20\% accuracy.

\section{Isotope shift calculation in ${\rm \bf Fe~II}$ and ${\rm \bf Fe~I}$}
\label{ms_in_FeI_and_II}

\subsection{Method}
It is known that the energy levels of two isotopes of any element are shifted relative
to each other. Total isotope shift (IS) is usually divided into mass shift and field shift. The former
is due to nuclear recoil and the latter is caused by the finite size of the nuclear
charge distribution. For light elements the field shift is much smaller than the
mass shift and we neglect it in our consideration.

In relativistic approximation the mass-shift operator can be represented
in the form of expansion in $\alpha Z$. In Ref.~\cite{ShaArt94} it was
obtained the following expression for this operator involving first
two terms of the expansion over $\alpha Z$
\begin{eqnarray}
\label{ms_rel_no_sym}
 H_\mathrm{MS} = \frac{1}{2M}\sum_{i,k}
 \left( \bm{p}_i \bm{p}_k  - \frac{\alpha Z}{r_i}
 \left[ \bm{\alpha}_i+
 (\bm{\alpha}_i{\bm{n}_i}){\bm{n}_i} \right] \bm{p}_k \right),
\end{eqnarray}
where $M$ is the nuclear mass, ${\bm p_i}$ is the momentum operator,
$\bm{n}_i = \bm{r}_i/r_i$, and ${\bm \alpha_i}$ is the Dirac matrice
of the $i$th electron.

The~\eref{ms_rel_no_sym} can be symmetrized over
variables of $i$th and $k$th electrons and written in the following form~\cite{KorKoz07}
\begin{eqnarray}
\label{ms_rel}
 H_\mathrm{MS} &=& \frac{1}{2M} \sum_{i,k}
 \left(  \bm{p}
 - \frac{\alpha Z}{2r}
 \left[\bm{\alpha}+
 (\bm{\alpha}\cdot{\bm{n}}){\bm{n}}\right]\right)_i
 \nonumber \\
&& \quad \quad \,\,\,\, \times
\left(  \bm{p}
 - \frac{\alpha Z}{2r}
 \left[\bm{\alpha}+
 (\bm{\alpha}\cdot{\bm{n}}){\bm{n}}\right]\right)_k.
\end{eqnarray}

This equation differs from~\eref{ms_rel_no_sym} only by the terms
$\sim (\alpha Z)^2$. \eref{ms_rel} is more convenient for CI calculations because
it is symmetric in electrons. This allows us to add it to the Coulomb integrals.

Using~\eref{ms_rel} we are able to calculate the isotope shift in
the frequency, $\delta \omega^{A,A'}$, of a transition between two isotopes with mass numbers
$A$ and $A'$. Neglecting the field shift the expression for $\delta \omega^{A,A'}$
can be written as
\begin{eqnarray}
\delta \omega^{A,A'} &=& \omega^{A} - \omega^{A'} \\ \nonumber
&\approx& (k_{\rm NMS} + k_{\rm SMS}) \left\{ \frac{1}{A} - \frac{1}{A'} \right\}.
\end{eqnarray}
The first term in curly brackets characterizes so-called normal mass shift (NMS),
while the second is the specific mass shift (SMS). In terms of~\eref{ms_rel} the
NMS term is determined by the expression with $i=k$ and the SMS term is determined
by the expression with $i \neq k$.

It is worth noting that the use of relativistic MS operator
is very important for calculating IS for the transitions between the FS
energy levels of the ground multiplet. An account for the relativistic corrections
changes drastically (up to change of the sign) the values of the IS obtained with
the nonrelativistic MS operator.

For transitions from the states of other multiplets to the ground state
the IS is less sensitive to the relativistic corrections.
In particular, for such transitions the coefficient $k_{\rm NMS}$ can be found
with a good accuracy from its nonrelativistic
expression $k_{\rm NMS} = -\omega/1823$. The value 1823 is the ratio of the atomic
mass unit to the electron mass.

Technically the isotope shift of an energy level can be found as follows.
The operator $H_{\rm MS}$ can be added to the many-body Hamiltonian $H$ with
an arbitrary coefficient $\lambda$:
\begin{equation}
H_\lambda = H + \lambda H_{\rm MS}.
\end{equation}

When the eigenvalue $E_\lambda$ of the Hamiltonian $H_\lambda$ is found, the IS correction to the
energy can be obtained as
\begin{equation}
\Delta E = \left. \frac{dE_\lambda}{d\lambda} \right|_{\lambda=0}
         \approx \frac{E_{+\lambda} - E_{-\lambda}}{2\lambda}.
\end{equation}
The parameter $\lambda$ should be chosen from the conditions of numerical
stability and smallness of the nonlinear terms. In our calculations
$\lambda/(2M)$ was put to be 0.001.

\subsection{Results for ${\rm \bf Fe~I}$}

We have carried out calculations of the isotope shifts for the transitions
from the even-parity states of the ground multiplet to the ground state
and from the odd-parity states of the $3d^6 4s4p$ and $3d^7 4p$ configurations
to the ground state using the [$4spdf$] basis set.
In the first case for calculation of both $k_{\rm SMS}$ and $k_{\rm NMS}$ we
used the relativistic MS operator given by~\eref{ms_rel}. In the second case
the specific mass shift $k_{\rm SMS}$ was calculated
in the relativistic approximation, while the normal mass shift was obtained from
the simple formula $k_{\rm NMS} = -\omega/1823$ with use of the experimental frequencies.

\begin{table}[bt]
\caption{Fe~\textsc{i}. Experimental and theoretical transition frequencies $\omega$
of the even- and odd-parity states respective to the ground state (in cm$^{-1}$),
$k_{\rm SMS}$ (in GHz) and $k_{\rm NMS}$ (in GHz) are presented.
The values are obtained for the [$4\!spdf$] basis set.}

\label{Tab_IS_FeI}

\begin{ruledtabular}
\begin{tabular}{rlcccc}
Config. & Level & \multicolumn{1}{c}{$\omega_{\rm exper}$\footnotemark[1]}
& \multicolumn{1}{c}{$\omega_{\rm theor}$}
& \multicolumn{1}{c}{$k_{\rm SMS}$} & \multicolumn{1}{c}{$k_{\rm NMS}$} \\[1mm]
\hline
$3d^6 4s^2$ & $a\,^5\!D^4$   & 0     & 0       & 0     &  0   \\
            & $a\,^5\!D^3$   & 416   & 409     & 6     & -3   \\
            & $a\,^5\!D^2$   & 704   & 697     & 10    & -5   \\
            & $a\,^5\!D^1$   & 888   & 883     & 12    & -6   \\[1.5mm]
$3d^6 4s4p$ & $z\,^7\!D^o_5$ & 19351 & 20550   & 557   & 318  \\
            & $z\,^7\!D^o_4$ & 19562 & 20759   & 562   & 322  \\
            & $z\,^7\!D^o_3$ & 19757 & 20972   & 565   & 325  \\[0.8mm]
            & $z\,^7\!F^o_5$ & 22846 & 22190   & 610   & 376  \\
            & $z\,^7\!F^o_4$ & 22997 & 22364   & 614   & 378  \\
            & $z\,^7\!F^o_3$ & 23111 & 22498   & 616   & 380 \\[0.8mm]
            & $z\,^7\!P^o_4$ & 23711 & 22727   & 650   & 390  \\
            & $z\,^7\!P^o_3$ & 24181 & 23199   & 658   & 398  \\[0.8mm]
            & $z\,^5\!D^o_4$ & 25900 & 25374   & 766   & 426  \\
            & $z\,^5\!D^o_3$ & 26140 & 25617   & 769   & 430 \\[0.8mm]
            & $z\,^5\!F^o_5$ & 26875 & 25885   & 734   & 442  \\
            & $z\,^5\!F^o_4$ & 27167 & 26189   & 743   & 447 \\
            & $z\,^5\!F^o_3$ & 27395 & 26416   & 748   & 450 \\[0.8mm]
            & $z\,^5\!P^o_3$ & 29056 & 27059   & 792   & 478  \\[0.8mm]
            & $z\,^3\!F^o_4$ & 31307 & 30147   & 981   & 515 \\
            & $z\,^3\!F^o_3$ & 31805 & 30624   & 988   & 523 \\[0.8mm]
            & $z\,^3\!D^o_3$ & 31323 & 29910   & 1016  & 515 \\[0.8mm]
            & $y\,^5\!P^o_3$ & 36767 & 38430   & 471   & 605 \\[0.8mm]
            & $x\,^5\!D^o_4$ & 39626 & 40609   & 2326  & 652 \\
            & $x\,^5\!D^o_3$ & 39970 & 41034   &       & 657 \\[0.8mm]
            & $x\,^5\!F^o_5$ & 40257 & 39620   & 1749  & 662 \\
            & $x\,^5\!F^o_4$ & 40594 & 39982   & 1810  & 668 \\
            & $x\,^5\!F^o_3$ & 40842 & 40258   &       & 672 \\[1.5mm]
$3d^7 4p$   & $y\,^5\!D^o_4$ & 33096 & 35157   & 1568  & 544 \\
            & $y\,^5\!D^o_3$ & 33507 & 35520   & 1529  & 551 \\[0.8mm]
            & $y\,^5\!F^o_5$ & 33695 & 36251   & 2287  & 554 \\
            & $y\,^5\!F^o_4$ & 34040 & 36571   & 2279  & 560 \\
            & $y\,^5\!F^o_3$ & 34329 & 36837   & 2257  & 565 \\[0.8mm]
            & $z\,^5\!G^o_5$ & 34782 & 38100   & 3508  & 572 \\
            & $z\,^5\!G^o_4$ & 35257 & 38499   & 3499  & 580 \\
            & $z\,^5\!G^o_3$ & 35612 & 38822   & 3506  & 586 \\[0.8mm]
            & $z\,^3\!G^o_5$ & 35379 & 38578   & 3514  & 582 \\
            & $z\,^3\!G^o_4$ & 35768 & 38913   & 3549  & 588 \\
            & $z\,^3\!G^o_3$ & 36079 & 39271   & 3531  & 593 \\[0.8mm]
            & $y\,^3\!F^o_4$ & 36686 & 39116   & 3521  & 603 \\
            & $y\,^3\!F^o_3$ & 37162 & 39507   & 3535  & 611 \\[0.8mm]
            & $y\,^3\!D^o_3$ & 38175 & 39933   & 3456  & 628 \\
\end{tabular}
\end{ruledtabular}
\footnotemark[1]{NIST, Ref.~\cite{NIST}}.
\end{table}

\tref{Tab_IS_FeI} presents the results obtained for $k_{\rm SMS}$ and $k_{\rm NMS}$.
Analyzing these results we see that the total mass shift
$k_{\rm MS} = k_{\rm SMS} + k_{\rm NMS}$ is small for the transitions from
the fine-structure components of the ground multiplet to the ground state.
Note that it is essential to account for both terms in~\eref{ms_rel_no_sym} for calculations
of these quantities. The contribution of the second term in~\eref{ms_rel_no_sym} to $k_{\rm SMS}$
is comparable to the contribution from the main term of the mass shift
operator. In contrast, relativistic corrections to the transition isotope mass shifts
of the odd-parity states are small and, in principle, can be neglected.

As it was shown in Ref.~\cite{BerFlaKoz08} the isotope shifts in Ti~\textsc{ii} for
certain levels are strongly influenced by core-valence correlations disregarded in our
approach. They contribute to $k_{\rm SMS}$ at the level of 50\% or even more.
Ti~\textsc{ii}, as well as Fe~\textsc{i}, is an element with open $d$ shell.
Its main configuration is $3d^2 4s$. Of course, the core of Ti~\textsc{ii} is less
rigid than that of Fe~\textsc{i} and, respectively, the role of core-valence correlations
should be greater. Nevertheless, we expect that a treatment of the core-valence
correlations will lead to more significant changes of $k_{\rm SMS}$ than inclusion of
the relativistic corrections to the interaction between valence electrons.

\subsection{Results for ${\rm \bf Fe~II}$}
 The method of calculation of different properties
of Fe~\textsc{ii} is similar to that for the neutral iron. The results of
calculation of the $q$ factors for a number of the odd-parity states were
reported in Refs.~\cite{PorKosTup07,DzuFlaKoz02}. All details of the method of calculations
can be found in~\cite{PorKosTup07}. Here we only briefly recapitulate its main features.

At first we solved the Dirac-Fock equations
to find core orbitals $1s$, ..., $3p_{3/2}$ and valence orbitals
$3d_{3/2}$, $3d_{5/2}$, $4p_{1/2}$, $4p_{3/2}$. Then we added virtual orbitals,
which were constructed using the procedure described in section~\ref{calc_FeI}.
The basis set used for seven-electron CI calculations included $s$, $p$, $d$,
and $f$ orbitals with principle quantum number $n\leq7$ designated as [$7spdf$].
Configuration space was formed by SD excitations from the configurations
$3d^6 4s$, $3d^6 4p$, and $3d^5 4s4p$.

The method of calculations of the IS remains the same as for Fe~\textsc{i}.
The results of the seven-electron CI calculation of the IS in Fe~\textsc{ii}
are listed in~\tref{Tab_IS_FeII}.
\begin{table}[tb!]
\caption{Fe~\textsc{ii}. Experimental and theoretical transition frequencies $\omega$
of the even- and odd-parity states respective to the ground state (in cm$^{-1}$),
$k_{\rm SMS}$ (in GHz) and $k_{\rm NMS}$ (in GHz) are presented.
$k_{\rm MS} = k_{\rm SMS} + k_{\rm NMS}$.
The values are obtained for the [$7\!spdf$] basis set.}

\label{Tab_IS_FeII}

\begin{ruledtabular}
\begin{tabular}{llcccccc}
Config. & Level & \multicolumn{1}{c}{$\omega_{\rm exper}$\footnotemark[1]}
& \multicolumn{1}{c}{$\omega_{\rm theor}$}
& \multicolumn{1}{c}{$k_{\rm SMS}$} & \multicolumn{1}{c}{$k_{\rm NMS}$}
& \multicolumn{1}{c}{$k_{\rm MS}$} & \multicolumn{1}{c}{Ref.~\cite{KozKorBer04}} \\[1mm]
\hline
$3d^6 4s$   & $^6\!D_{9/2}$   &    0  &    0  &    0  &  0  &    0 &            \\
            & $^6\!D_{7/2}$   &   385 &   375 &    7  & -3  &    4 &            \\
            & $^6\!D_{5/2}$   &   668 &   653 &   11  & -4  &    7 &            \\
            & $^6\!D_{3/2}$   &   863 &   846 &   14  & -5  &    9 &            \\
            & $^6\!D_{1/2}$   &   977 &   960 &   15  & -6  &    9 &            \\[1.5mm]
$3d^6 4p$   & $^6\!D^o_{9/2}$ & 38459 & 37373 &  572  & 632 & 1204 & 1800(600)  \\
            & $^6\!D^o_{7/2}$ & 38660 & 37573 &  576  & 636 & 1212 & 1800(600)  \\[0.8mm]
            & $^6\!F^o_{11/2}$& 41968 & 41097 &  753  & 690 & 1443 & 1900(600)  \\
            & $^6\!F^o_{9/2}$ & 42115 & 41247 &  759  & 693 & 1452 & 1900(600)  \\
            & $^6\!F^o_{7/2}$ & 42237 & 41370 &  762  & 695 & 1457 &            \\[0.8mm]
            & $^6\!P^o_{7/2}$ & 42658 & 41760 &  468  & 702 & 1170 & 1800(600)  \\[0.8mm]
            & $^6\!D^o_{7/2}$ & 38660 & 37573 &  576  & 636 & 1212 &            \\[0.8mm]
            & $^6\!F^o_{7/2}$ & 42237 & 41370 &  762  & 695 & 1457 &            \\[0.8mm]
            & $^6\!P^o_{7/2}$ & 42658 & 41760 &  468  & 702 & 1170 &            \\[0.8mm]
            & $^4\!F^o_{7/2}$ & 44754 & 44044 &  737  & 736 & 1473 & 2010(1200) \\[0.8mm]
            & $^4\!D^o_{7/2}$ & 44447 & 44270 &  728  & 731 & 1459 &            \\[0.8mm]
            & $^4\!G^o_{7/2}$ & 60957 & 62766 &  626  &1002 & 1628 &            \\[0.8mm]
            & $^4\!H^o_{7/2}$ & 61157 & 62894 &  947  &1006 & 1953 &            \\[0.8mm]
            & $^4\!F^o_{7/2}$ & 62066 & 64017 &  305  &1021 & 1326 &            \\[0.8mm]
            & $^2\!G^o_{7/2}$ & 62322 & 64217 &  526  &1025 & 1551 &            \\[1.5mm]
$3d^5 4s4p$ & $^8\!P^o_{7/2}$ & 52583 & 49115 &-3164  & 865 &-2299 &            \\[0.8mm]
            & $^6\!P^o_{7/2}$ & 62172 & 59245 &-3074  &1022 &-2052 &-2010(1200) \\
\end{tabular}
\end{ruledtabular}
\footnotemark[1]{NIST, Ref.~\cite{NIST}}.
\end{table}
To the best of our knowledge the only calculation of the IS for certain transitions
in Fe~\textsc{ii} was carried out in~\cite{KozKorBer04}. We see reasonable agreement between
the results obtained in this work with the values found in Ref.~\cite{KozKorBer04}.

\section{Conclusion}
\label{conclusions}

We have calculated relativistic frequency shifts ($q$ factors) for a
number of transitions from the even- and odd-parity states of Fe~\textsc{i} to
the ground state. The calculations were carried out for the three basis
sets [$4spdf$], [$5spdf$], and [$6spdf$]. Comparing the results
obtained for these basis sets we could estimate the accuracy of
the $q$ factors for the odd-parity
energy levels listed in Tables~\ref{Tab_od_wgq} and \ref{Tab_d7p}
at the level of 20\%. The accuracy of the $q$ factors for the
transitions from the even-parity states of the ground multiplet to the
ground state is significantly higher. We estimate it at the level of
a few percent.

The certain odd-parity states are astrophysically
interesting because they were observed in quasar absorption
spectra. Due to strong configuration interaction
the magnitudes of the $q$ factors vary significantly between the states. This
makes Fe~\textsc{i} a very attractive candidate to search for
hypothetical $\alpha$ variation.

We also computed the mass isotope shifts for
Fe~\textsc{i} and Fe~\textsc{ii} for the [$4spdf$] basis set.
Comparing the results obtained for Fe~\textsc{ii} with the values
of~\cite{KozKorBer04} we see a reasonable agreement between them.
To the best of our knowledge, the isotope shifts in Fe~\textsc{i}
were calculated for the first time.

\section{acknowledgments}

We would like to thank V.~Dzuba, S.~Levshakov, and P.~Molaro who brought our attention
to this problem. This work was supported in part by the Russian Foundation for Basic
Research under Grants No. 07-02-00210-a and No. 08-02-00460-a,
and by DFG Grants No. SFB 676 Teilprojekt C4 and No. RE 353/48-1.

\bibliographystyle{apsrev}%


\begin{thebibliography}{26}
\expandafter\ifx\csname natexlab\endcsname\relax\def\natexlab#1{#1}\fi
\expandafter\ifx\csname bibnamefont\endcsname\relax
  \def\bibnamefont#1{#1}\fi
\expandafter\ifx\csname bibfnamefont\endcsname\relax
  \def\bibfnamefont#1{#1}\fi
\expandafter\ifx\csname citenamefont\endcsname\relax
  \def\citenamefont#1{#1}\fi
\expandafter\ifx\csname url\endcsname\relax
  \def\url#1{\texttt{#1}}\fi
\expandafter\ifx\csname urlprefix\endcsname\relax\def\urlprefix{URL }\fi
\providecommand{\bibinfo}[2]{#2}
\providecommand{\eprint}[2][]{\url{#2}}

\bibitem[{\citenamefont{{\rm Carc\'{i}a-Berro} et~al.}(2007)\citenamefont{{\rm
  Carc\'{i}a-Berro}, Isern, and Kubyshin}}]{GarIseKub07}
\bibinfo{author}{\bibfnamefont{E.}~\bibnamefont{{\rm Carc\'{i}a-Berro}}},
  \bibinfo{author}{\bibfnamefont{J.}~\bibnamefont{Isern}}, \bibnamefont{and}
  \bibinfo{author}{\bibfnamefont{Y.~A.} \bibnamefont{Kubyshin}},
  \bibinfo{journal}{Astron. Astrophys. Rev.} \textbf{\bibinfo{volume}{14}},
  \bibinfo{pages}{113} (\bibinfo{year}{2007}).

\bibitem[{\citenamefont{Copeland et~al.}(2006)\citenamefont{Copeland, Sami, and
  Tsujikawa}}]{CopSamTsu06}
\bibinfo{author}{\bibfnamefont{E.~J.} \bibnamefont{Copeland}},
  \bibinfo{author}{\bibfnamefont{M.}~\bibnamefont{Sami}}, \bibnamefont{and}
  \bibinfo{author}{\bibfnamefont{S.}~\bibnamefont{Tsujikawa}},
  \bibinfo{journal}{Int. J. Mod. Phys. D} \textbf{\bibinfo{volume}{15}},
  \bibinfo{pages}{1753} (\bibinfo{year}{2006}).

\bibitem[{\citenamefont{Murphy et~al.}(2003)\citenamefont{Murphy, Webb, and
  Flambaum}}]{MurWebFla03}
\bibinfo{author}{\bibfnamefont{M.~T.} \bibnamefont{Murphy}},
  \bibinfo{author}{\bibfnamefont{J.~K.} \bibnamefont{Webb}}, \bibnamefont{and}
  \bibinfo{author}{\bibfnamefont{V.~V.} \bibnamefont{Flambaum}},
  \bibinfo{journal}{Mon. Not. R. Astron. Soc.} \textbf{\bibinfo{volume}{345}},
  \bibinfo{pages}{609} (\bibinfo{year}{2003}).

\bibitem[{\citenamefont{Quast et~al.}(2004)\citenamefont{Quast, Reimers, and
  Levshakov}}]{QRL04}
\bibinfo{author}{\bibfnamefont{R.}~\bibnamefont{Quast}},
  \bibinfo{author}{\bibfnamefont{D.}~\bibnamefont{Reimers}}, \bibnamefont{and}
  \bibinfo{author}{\bibfnamefont{S.~A.} \bibnamefont{Levshakov}},
  \bibinfo{journal}{Astron. \& Astrophys.} \textbf{\bibinfo{volume}{414}},
  \bibinfo{pages}{L7} (\bibinfo{year}{2004}).

\bibitem[{\citenamefont{Srianand et~al.}(2004)\citenamefont{Srianand, Chand,
  Petitjean, and Aracil}}]{SCP04}
\bibinfo{author}{\bibfnamefont{R.}~\bibnamefont{Srianand}},
  \bibinfo{author}{\bibfnamefont{H.}~\bibnamefont{Chand}},
  \bibinfo{author}{\bibfnamefont{P.}~\bibnamefont{Petitjean}},
  \bibnamefont{and} \bibinfo{author}{\bibfnamefont{B.}~\bibnamefont{Aracil}},
  \bibinfo{journal}{Phys. Rev. Lett.} \textbf{\bibinfo{volume}{92}},
  \bibinfo{pages}{121302} (\bibinfo{year}{2004}).

\bibitem[{\citenamefont{Levshakov et~al.}(2007)\citenamefont{Levshakov, Molaro,
  Lopez et~al.}}]{LML07}
\bibinfo{author}{\bibfnamefont{S.~A.} \bibnamefont{Levshakov}},
  \bibinfo{author}{\bibfnamefont{P.}~\bibnamefont{Molaro}},
  \bibinfo{author}{\bibfnamefont{S.}~\bibnamefont{Lopez}},
  \bibnamefont{et~al.}, \bibinfo{journal}{Astron. \& Astrophys.}
  \textbf{\bibinfo{volume}{466}}, \bibinfo{pages}{1077} (\bibinfo{year}{2007}).

\bibitem[{\citenamefont{{\rm D'Odorico}}(2007)}]{Dod07}
\bibinfo{author}{\bibfnamefont{V.}~\bibnamefont{{\rm D'Odorico}}},
  \bibinfo{journal}{Astron. Astrophys.} \textbf{\bibinfo{volume}{470}},
  \bibinfo{pages}{523} (\bibinfo{year}{2007}).

\bibitem[{\citenamefont{Quast et~al.}(2008)\citenamefont{Quast, Reimers, and
  Baade}}]{QuaReiBaa08}
\bibinfo{author}{\bibfnamefont{R.}~\bibnamefont{Quast}},
  \bibinfo{author}{\bibfnamefont{D.}~\bibnamefont{Reimers}}, \bibnamefont{and}
  \bibinfo{author}{\bibfnamefont{R.}~\bibnamefont{Baade}},
  \bibinfo{journal}{Astron. Astrophys.} \textbf{\bibinfo{volume}{477}},
  \bibinfo{pages}{443} (\bibinfo{year}{2008}).

\bibitem[{ML0()}]{ML07}
\bibinfo{note}{P. Molaro and S. A. Levshakov, private communication (2007).}

\bibitem[{\citenamefont{Levshakov}(1994)}]{Lev94}
\bibinfo{author}{\bibfnamefont{S.~A.} \bibnamefont{Levshakov}},
  \bibinfo{journal}{Mon. Not. R. Astron. Soc.} \textbf{\bibinfo{volume}{269}},
  \bibinfo{pages}{339} (\bibinfo{year}{1994}).

\bibitem[{\citenamefont{Varshalovich et~al.}(2001)\citenamefont{Varshalovich,
  Potekhin, and Ivanchik}}]{VPI01}
\bibinfo{author}{\bibfnamefont{D.~A.} \bibnamefont{Varshalovich}},
  \bibinfo{author}{\bibfnamefont{A.~Y.} \bibnamefont{Potekhin}},
  \bibnamefont{and} \bibinfo{author}{\bibfnamefont{A.~V.}
  \bibnamefont{Ivanchik}}, \bibinfo{journal}{Phys. Scripta}
  \textbf{\bibinfo{volume}{T95}}, \bibinfo{pages}{76} (\bibinfo{year}{2001}).

\bibitem[{\citenamefont{Kozlov et~al.}(2004)\citenamefont{Kozlov, Korol,
  Berengut, Dzuba, and Flambaum}}]{KozKorBer04}
\bibinfo{author}{\bibfnamefont{M.~G.} \bibnamefont{Kozlov}},
  \bibinfo{author}{\bibfnamefont{V.~A.} \bibnamefont{Korol}},
  \bibinfo{author}{\bibfnamefont{J.~C.} \bibnamefont{Berengut}},
  \bibinfo{author}{\bibfnamefont{V.~A.} \bibnamefont{Dzuba}}, \bibnamefont{and}
  \bibinfo{author}{\bibfnamefont{V.~V.} \bibnamefont{Flambaum}},
  \bibinfo{journal}{Phys. Rev. A} \textbf{\bibinfo{volume}{70}},
  \bibinfo{pages}{062108} (\bibinfo{year}{2004}).

\bibitem[{\citenamefont{Korol and Kozlov}(2007)}]{KorKoz07}
\bibinfo{author}{\bibfnamefont{V.~A.} \bibnamefont{Korol}} \bibnamefont{and}
  \bibinfo{author}{\bibfnamefont{M.~G.} \bibnamefont{Kozlov}},
  \bibinfo{journal}{Phys. Rev. A} \textbf{\bibinfo{volume}{76}},
  \bibinfo{pages}{022103} (\bibinfo{year}{2007}).

\bibitem[{\citenamefont{Dzuba and Flambaum}(2008)}]{DzuFla08}
\bibinfo{author}{\bibfnamefont{V.~A.} \bibnamefont{Dzuba}} \bibnamefont{and}
  \bibinfo{author}{\bibfnamefont{V.~V.} \bibnamefont{Flambaum}},
  \bibinfo{journal}{Phys. Rev. A} \textbf{\bibinfo{volume}{77}},
  \bibinfo{pages}{012514} (\bibinfo{year}{2008}).

\bibitem[{\citenamefont{Bogdanovich}(1991)}]{Bog91}
\bibinfo{author}{\bibfnamefont{P.}~\bibnamefont{Bogdanovich}},
  \bibinfo{journal}{Lith. Phys. J.} \textbf{\bibinfo{volume}{31}},
  \bibinfo{pages}{79} (\bibinfo{year}{1991}).

\bibitem[{\citenamefont{Kozlov et~al.}(1996)\citenamefont{Kozlov, Porsev, and
  Flambaum}}]{KozPorFla96}
\bibinfo{author}{\bibfnamefont{M.~G.} \bibnamefont{Kozlov}},
  \bibinfo{author}{\bibfnamefont{S.~G.} \bibnamefont{Porsev}},
  \bibnamefont{and} \bibinfo{author}{\bibfnamefont{V.~V.}
  \bibnamefont{Flambaum}}, \bibinfo{journal}{J. \ Phys. \ B}
  \textbf{\bibinfo{volume}{29}}, \bibinfo{pages}{689} (\bibinfo{year}{1996}).

\bibitem[{\citenamefont{Porsev et~al.}(2007)\citenamefont{Porsev, Koshelev,
  Tupitsyn, Kozlov, Reimers, and Levshakov}}]{PorKosTup07}
\bibinfo{author}{\bibfnamefont{S.~G.} \bibnamefont{Porsev}},
  \bibinfo{author}{\bibfnamefont{K.~V.} \bibnamefont{Koshelev}},
  \bibinfo{author}{\bibfnamefont{I.~I.} \bibnamefont{Tupitsyn}},
  \bibinfo{author}{\bibfnamefont{M.~G.} \bibnamefont{Kozlov}},
  \bibinfo{author}{\bibfnamefont{D.}~\bibnamefont{Reimers}}, \bibnamefont{and}
  \bibinfo{author}{\bibfnamefont{S.~A.} \bibnamefont{Levshakov}},
  \bibinfo{journal}{Phys. Rev. A} \textbf{\bibinfo{volume}{76}},
  \bibinfo{pages}{052507} (\bibinfo{year}{2007}).

\bibitem[{\citenamefont{Greenberg et~al.}(1977)\citenamefont{Greenberg, Dyal,
  and Geballe}}]{GreDyaGeb77}
\bibinfo{author}{\bibfnamefont{L.~T.} \bibnamefont{Greenberg}},
  \bibinfo{author}{\bibfnamefont{P.}~\bibnamefont{Dyal}}, \bibnamefont{and}
  \bibinfo{author}{\bibfnamefont{T.~R.} \bibnamefont{Geballe}},
  \bibinfo{journal}{Astrophys. J.} \textbf{\bibinfo{volume}{213}},
  \bibinfo{pages}{L71} (\bibinfo{year}{1977}).

\bibitem[{\citenamefont{Moorwood et~al.}(1980)\citenamefont{Moorwood, Salinari,
  Furniss, Jennings, and King}}]{MooSalFur80}
\bibinfo{author}{\bibfnamefont{A.~{\rm F. M}.} \bibnamefont{Moorwood}},
  \bibinfo{author}{\bibfnamefont{P.}~\bibnamefont{Salinari}},
  \bibinfo{author}{\bibfnamefont{I.}~\bibnamefont{Furniss}},
  \bibinfo{author}{\bibfnamefont{R.~E.} \bibnamefont{Jennings}},
  \bibnamefont{and} \bibinfo{author}{\bibfnamefont{K.~J.} \bibnamefont{King}},
  \bibinfo{journal}{Astron. Astrophys.} \textbf{\bibinfo{volume}{90}},
  \bibinfo{pages}{304} (\bibinfo{year}{1980}).

\bibitem[{\citenamefont{Wei{\ss} et~al.}(2003)\citenamefont{Wei{\ss}, Henkel,
  Downes, and Walter}}]{WeiHenDow03}
\bibinfo{author}{\bibfnamefont{A.}~\bibnamefont{Wei{\ss}}},
  \bibinfo{author}{\bibfnamefont{C.}~\bibnamefont{Henkel}},
  \bibinfo{author}{\bibfnamefont{D.}~\bibnamefont{Downes}}, \bibnamefont{and}
  \bibinfo{author}{\bibfnamefont{F.}~\bibnamefont{Walter}},
  \bibinfo{journal}{Astron. Astrophys.} \textbf{\bibinfo{volume}{409}},
  \bibinfo{pages}{L41} (\bibinfo{year}{2003}).

\bibitem[{NIS()}]{NIST}
\emph{\bibinfo{title}{{\rm NIST,} {A}tomic {S}pectra {D}atabase}},
  \urlprefix\url{http://physics.nist.gov/cgi-bin/AtData/main_asd}.

\bibitem[{\citenamefont{Kozlov et~al.}(2008)\citenamefont{Kozlov, Porsev,
  Levshakov, Reimers, and Molaro}}]{KozPorLev08}
\bibinfo{author}{\bibfnamefont{M.~G.} \bibnamefont{Kozlov}},
  \bibinfo{author}{\bibfnamefont{S.~G.} \bibnamefont{Porsev}},
  \bibinfo{author}{\bibfnamefont{S.~A.} \bibnamefont{Levshakov}},
  \bibinfo{author}{\bibfnamefont{D.}~\bibnamefont{Reimers}}, \bibnamefont{and}
  \bibinfo{author}{\bibfnamefont{P.}~\bibnamefont{Molaro}},
  \bibinfo{journal}{Phys. Rev. A} \textbf{\bibinfo{volume}{77}},
  \bibinfo{pages}{032119} (\bibinfo{year}{2008}).

\bibitem[{\citenamefont{Dzuba et~al.}(1999)\citenamefont{Dzuba, Flambaum, and
  Webb}}]{DzuFlaWeb99}
\bibinfo{author}{\bibfnamefont{V.~A.} \bibnamefont{Dzuba}},
  \bibinfo{author}{\bibfnamefont{V.~V.} \bibnamefont{Flambaum}},
  \bibnamefont{and} \bibinfo{author}{\bibfnamefont{J.~K.} \bibnamefont{Webb}},
  \bibinfo{journal}{Phys. Rev. A} \textbf{\bibinfo{volume}{59}},
  \bibinfo{pages}{230} (\bibinfo{year}{1999}).

\bibitem[{\citenamefont{Shabaev and Artemyev}(1994)}]{ShaArt94}
\bibinfo{author}{\bibfnamefont{V.~M.} \bibnamefont{Shabaev}} \bibnamefont{and}
  \bibinfo{author}{\bibfnamefont{A.~N.} \bibnamefont{Artemyev}},
  \bibinfo{journal}{J. Phys. B} \textbf{\bibinfo{volume}{27}},
  \bibinfo{pages}{1307} (\bibinfo{year}{1994}).

\bibitem[{\citenamefont{Berengut et~al.}(2008)\citenamefont{Berengut, Flambaum,
  and Kozlov}}]{BerFlaKoz08}
\bibinfo{author}{\bibfnamefont{J.~C.} \bibnamefont{Berengut}},
  \bibinfo{author}{\bibfnamefont{V.~V.} \bibnamefont{Flambaum}},
  \bibnamefont{and} \bibinfo{author}{\bibfnamefont{M.~G.}
  \bibnamefont{Kozlov}}, \bibinfo{journal}{J. Phys. B}
  \textbf{\bibinfo{volume}{41}}, \bibinfo{pages}{235702}
  (\bibinfo{year}{2008}).

\bibitem[{\citenamefont{Dzuba et~al.}(2002)\citenamefont{Dzuba, Flambaum,
  Kozlov, and Marchenko}}]{DzuFlaKoz02}
\bibinfo{author}{\bibfnamefont{V.~A.} \bibnamefont{Dzuba}},
  \bibinfo{author}{\bibfnamefont{V.~V.} \bibnamefont{Flambaum}},
  \bibinfo{author}{\bibfnamefont{M.~G.} \bibnamefont{Kozlov}},
  \bibnamefont{and}
  \bibinfo{author}{\bibfnamefont{M.}~\bibnamefont{Marchenko}},
  \bibinfo{journal}{Phys. Rev. A} \textbf{\bibinfo{volume}{66}},
  \bibinfo{pages}{022501} (\bibinfo{year}{2002}).

\end{thebibliography}

\end{document}